\documentclass[runningheads]{llncs}

\author{Karen Wintersperger\inst{1}\orcidID{0000-0002-2181-1860} \and
Hila Safi\inst{1,2}\and\\
Wolfgang Mauerer\inst{2,1}\orcidID{0000-0002-9765-8313}}
\authorrunning{Wintersperger et al.}
\institute{Siemens AG, Corporate Technology,\\ Otto-Hahn-Ring 6,
81739 München, Germany\\
\email{\{karen.wintersperger,hila.safi\}@siemens.com}
\and
Technical University of Applied Sciences Regensburg,\\
Galgenbergstraße 32, 93053 Regensburg, Germany\\
\email{wolfgang.mauerer@othr.de}}

\usepackage{cite}
\usepackage{amsmath,amssymb,amsfonts}
\usepackage{algorithmic}
\usepackage{graphicx}
\usepackage{textcomp}
\usepackage{xcolor}
\usepackage{booktabs}
\usepackage{xspace}
\usepackage{multirow}
\usepackage{braket}
\usepackage{subcaption}
\usepackage{csquotes}
\usepackage{siunitx}
\usepackage{hyperref}
\usepackage{tikz}
\usetikzlibrary{positioning, calc, arrows.meta} 
\usepackage{wrapfig}

\newcommand{\ie}{\emph{i.e.}\xspace}

\newcommand{\etal}{\emph{et al.}\xspace}
\newcommand{\bigO}[1]{\ensuremath{\mathcal{O}(#1)}}
\newcommand{\nsa}{\ensuremath{n_{\text{samp}}}}
\newcommand{\nit}{\ensuremath{n_{\text{iter}}}}
\newcommand{\tci}{\ensuremath{t_{\text{circ}}}}
\newcommand{\tco}{\ensuremath{t_{\text{comm}}}}
\newcommand{\tm}{\ensuremath{t_{\text{meas}}}}
\newcommand{\topt}{\ensuremath{t_{\text{opt}}}}
\newcommand{\CXgate}{\ensuremath{\text{C\raisebox{0.08em}{--}}\!X}\xspace}

\hypersetup{
    colorlinks,
    linkcolor={black},
    citecolor={black},
    urlcolor={blue}
}

\begin{document}

\title{QPU-System Co-Design for\\ Quantum HPC Accelerators\thanks{All authors acknowledge funding from the German Federal Ministry of Education and Research within the funding program quantum technologies---from basic research to market, contract number 13N16093.}}
\maketitle

\begin{abstract}

The use of quantum processing units (QPUs) promises speed-ups for solving computational
problems, but the quantum devices currently available possess only a very limited number
of qubits and suffer from considerable imperfections. One possibility to progress towards
practical utility is to use a co-design approach: 
Problem formulation and algorithm, but also the physical QPU properties
are tailored to the specific application. Since QPUs will likely be used as accelerators
for classical computers, details of systemic integration into existing architectures
are another lever to influence and improve the practical utility of QPUs.

In this work, we investigate the influence of different parameters on the runtime of
quantum programs on tailored hybrid CPU-QPU-systems. We study the influence of communication
times between CPU and QPU, how adapting QPU designs influences quantum and overall
execution performance, and how these factors interact. Using a simple model 
that allows for estimating which design choices should be subjected to optimisation for a
given task, we provide an intuition to the HPC community on potentials and limitations of
co-design approaches. We also discuss physical limitations for implementing the
proposed changes on real quantum hardware devices.


\end{abstract}

\section{Introduction} 


\looseness-1 Quantum computers available today are restricted in their performance by relatively
small ($\approx$ 50--100) number of quantum bits (qubits), and suffer from
various imperfections. Since they cannot yet implement error correction routinely,
they classify as NISQ (Noisy Intermediate Scale Quantum) hardware~\cite{preskill_2018},
whose capabilities are subject to ongoing exploration. 
Nevertheless, there is a growing interest to deploy NISQ devices in high-performance  
computing (HPC) scenarios.~\cite{Bayerstadler_2021}

Fitting complex problems to NISQ devices usually requires many simplifications.
Also, the mapping of the problem to a quantum model and the algorithm can be optimised
to reduce both, the number of necessary qubits, and the number of quantum operations~\cite{Kruger_2020}.
%
Properties of quantum algorithms strongly depend on specific QPUs. Many factors,
size (number of qubits), the geometric arrangement of and number of connections
between the qubits, as well as the specific errors and execution times of quantum
operations, influence quantum circuit execution. Prior to running a quantum circuit
on a quantum device, the circuit structure must be adjusted to its requirements,
which usually increases the number of quantum operations, and thus the circuit
depth. Co-designing QPUs by adapting them to specific problem classes is therefore a
promising approach.  

QPUs will likely be used as accelerators for classical computers, and require
integration with classical hardware. Moreover, hybrid algorithms that combine
classical and quantum operations are a commonly occurring pattern. Thus, the
interaction between quantum and classical devices needs to be taken into account
to estimate the performance of a quantum program in practice.

In this work, we investigate optimisation potentials for the co-design of such CPU-QPU-systems,
exemplified by a hybrid quantum algorithm used to solve the maximum cut (Max-Cut) problem. 
In detail, our contributions are as follows:
\begin{itemize}
\item We analyse the properties of compiled quantum circuits for various instances
of the Max-Cut problem when modifying geometric properties of the QPU, namely the
connectivity between qubits and the number of qubits.
\item We estimate the runtime on real quantum hardware based on simulations, and
investigate the overall runtime on QPU-systems including the communication
between classical and quantum machines, and other classical calculations 
required to execute and evaluate quantum circuits.
\item Based on these results, we give recommendations for the design of quantum hardware adapted to applications. 
\item We provide a self-contained replication package~\cite{mauerer_2022}
 for the simulations that is available at \url{https://github.com/lfd/arcs2022.git}.
\end{itemize}

The Max-Cut problem is an optimisation problem with applications
in network design, clustering or statistical physics. Considering an undirected graph
with a set of nodes $V$ and a set of edges $E$, a cut is defined as a partition of the
node set into two subsets. The Max-Cut problem seeks to find a partition such that the
number of edges connecting the two subsets is maximised. Many applications
for this primitive can be found in all areas of computing; for us, it suffices
to state that Max-Cut serves as typical representative for hard problems that
we will further motivate in Section~\ref{sec:background}.

There are different, yet equivalent models for quantum computation, such as
gate-based~\cite{deutsch_1989, barenco_1995}, measurement-based~\cite{briegel_2003},
adiabatic~\cite{farhi_2001}, or topological quantum
computing. In this work, we focus on gate-based quantum computation.
The Max-Cut problem is solved using the Quantum Approximate Optimisation Algorithm
(QAOA)~\cite{farhi_2014}, which is a widely used variational hybrid quantum algorithm
for solving combinatorial optimisation problems on NISQ hardware.


As a starting point for the investigation of quantum hardware designs,
IBM-Q devices are used, which are based on superconducting qubit
technology~\cite{huang_2020_SC_rev, kjaergaard_2020_SC_rev}. Superconducting qubits are one
of the most common and advanced quantum hardware platforms, used by many vendors such as 
\href{https://quantum-computing.ibm.com}{IBM},  \href{https://quantumai.google/}{Google} and
\href{https://qcs.rigetti.com/qpus}{Rigetti}. 
Since superconducting qubits are artificial quantum systems, they can in principle be designed at will. By now, several different types of superconducting qubits exist, and the technology is continuously
developed. However, this also means that no two qubits are completely identical, and
properties such as the gate fidelity, which is a measure for the quality of a quantum operation, differ for each qubit. 
The superconducting quantum devices of different vendors usually differ by the
geometric arrangement of qubits, and the number and structure of connections
between them. The IBM-Q topology that we use as basis for our considerations is
described in detail in Section~\ref{subsec:setup}.

\section{Quantum Max-Cut with QAOA}\label{sec:base_top}  
Before we discuss co-design optimisation possibilities, we need to set the stage
for the considered problem, and illustrate the solution algorithm. 

\subsection{The Quantum Approximate Optimisation Algorithm}\label{sec:qaoa_intro}
QAOA implements a quantum circuit consisting of $p \in \mathbb{N}$ layers of unitary operators (the elementary type of operation a quantum computer can effect on
qubits) whose properties are specified by a set of $2p$
parameters $\vec{\beta}, \vec{\gamma} \in\mathbb{R}^{p}$. The algorithm can
determine minima of objective functions specified in quadratic, unconstrained
binary form; these are specified such that the minimum solution corresponds
to a solution of a specific problem of interest. Using well-known
techniques from computer science, all problems in \(\mathbf{NP}\) can 
be reduced to \emph{Quadratic Unconstrained Binary Optimisation} (QUBO) form~\cite{lucas_2014}. Speedups of QAOA compared to
classical approaches are not yet fully understood~\cite{bharti_2022}; yet
the existence of a classical algorithm that efficiently samples the output distribution of
QAOA even for $p=1$ is impossible, given reasonable complexity-theoretic
assumptions~\cite{farhi_2016}.  This is seen as likely indicator for quantum advantage,
but theoretical or experimental progress is required to find utility on practical
problems.

\begin{figure}[htb]
\centering
  \includegraphics[width=0.7\textwidth]{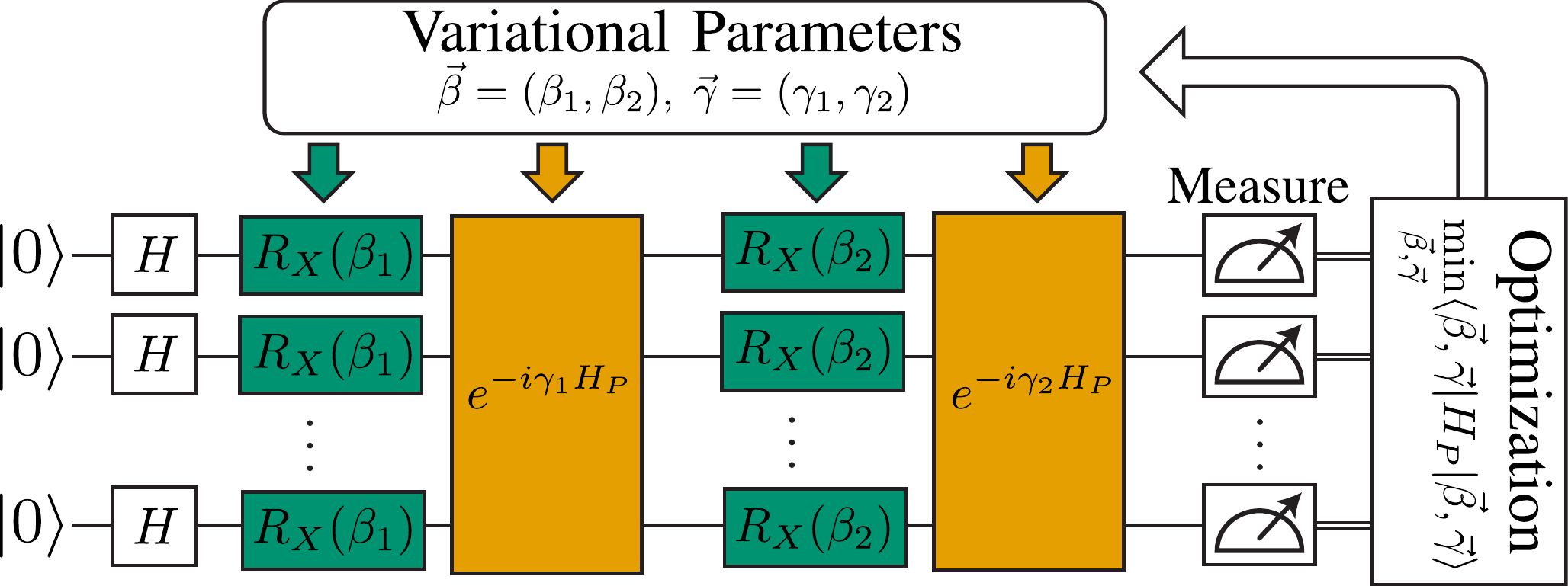}
  \caption{Sketch of the QAOA-circuit for $p=2$.}
  \label{fig:sketch_QAOA}
\end{figure}

Fig.~\ref{fig:sketch_QAOA} sketches the structure of the QAOA circuit for $p=2$: After applying the operators to a well-defined initial state, the expectation value of $H_P$, which
encodes the objective function, is
measured in the final state. Using a classical optimiser, the parameters of the circuit are changed with the goal of minimising the expectation value of $H_P$.
Each layer $i$ consists of two different kinds of unitaries: First, $U(\beta_i) = e^{-i \beta H_B}$ is applied, implementing the evolution under a so-called mixer Hamiltonian $H_B$. The mixer Hamiltonian is commonly chosen as a superposition of $X$-rotations applied to each qubit, thus consisting of a series of rotation gates $R_X(\beta_i)$.
The second part of each layer comprises $U(\gamma_i) = e^{-i \gamma H_P}$ consisting of single-qubit $Z$-rotations $R_Z(\gamma_i)$ 
and two-qubit rotation gates $R_{ZZ}(\gamma_i)$. 
The repeated application of several QAOA layers corresponds to the discretised time evolution governed by the Hamiltonians $H_P$ and $H_B$. It is known that
the quality of the approximation increases for a larger number of layers~\cite{farhi_2014}. The initial state of the QAOA algorithm is usually chosen as the ground state of $H_B$, in which each qubit is in an equal superposition of \(\ket{0}\) and \(\ket{1}\),  prepared using a layer of Hadamard gates $H$. 

To characterise the probability distribution of the final state depending on \(\vec{\beta}, \vec{\gamma}\) after each iteration of the optimiser, the quantum circuit is executed several times (known as \emph{sampling}), and the qubits are measured in the computational basis \(\{\ket{0}, \ket{1}\}\). The mean of the expectation values of $H_P$ for each measurement outcome is used as the objective function minimised by the classical optimiser. The optimal solution is then given by the state (or bit string) with the lowest energy expectation value taken from the probability distribution obtained for the final set of parameters. 

\subsection{Background on Max-Cut and QAOA}\label{sec:background}

Given a graph \(G=(V,E)\) consisting of a set of vertices \(V\) and a set of edges
\(E\subseteq V\times V\), the Max-Cut problem is a seminal graph-theoretic task that seeks
two subsets \(V_{0}, V_{1} \subseteq V\) such that \(V_{0}\cup V_{1} = V\) and \(V_{0}\cap
V_{1}=\varnothing\), and maximises the size \(|C|\) of the cut set
\(C = \{(u,v)\in E: u\in V_{0}, v\in V_{1}\}\) (as a decision problem, the k-Cut variant seeks
a cut with \(|C| = k\)). 
While the Max-Cut problem is very simple to formulate, it counts among the hardest
optimisation problems to solve~\cite{commander_2009}. It is textbook knowledge that the decision variant lies in
complexity class \(\mathbf{NP}\), while the optimisation variant is \(\mathbf{APX}\text{-hard}\),
which essentially means that any polynomial-time approximation algorithm can at best find
solutions whose approximation ratio (\ie, the size of the found cut size divided by the optimal
cut size) is bounded by a constant for general graphs.

 Classical approximation algorithms
and heuristics for Max-Cut, including formulations adapted to graphs with
specific properties, have been studied in the literature, and polynomial-time approximation
algorithms with non-trivial performance guarantees are known. In particular, the seminal
algorithm of Goemans and Williamson~\cite{goemans_1995} achieves an approximation ratio
of 87.86\% on generic graphs, while Khot~\etal discuss the optimality of inapproximability
results for the problem. Ever since Farhi~\etal~\cite{farhi_2014} found that QAOA
at circuit depth \(p=1\) (a) already achieves an approximation ratio of at least 69.25\% (for the
class of uniform, 3-regular graphs), yet (b) cannot be efficiently simulated by classical
algorithms, assuming the validity of widely accepted complexity-theoretic hypotheses, there
has been a steady interest in understanding properties and performance guarantees of QAOA
on the Max-Cut problem.

Fuchs~\etal~\cite{fuchs_2021} present an efficient encoding of (weighted) Max-Cut for QAOA,
while Wurz and Lykov~\cite{wurtz_2021a} discuss open conjectures regarding the quantum performance
of the problem. On the negative side from the QPU point of view, Marwaha~\cite{marwaha_2021} showed
that classical algorithms outperform QAOA with \(p=2\) with a certain type of graphs. On the positive
side for quantum algorithms, Wurtz and Love~\cite{wurtz_2021b} give performance guarantees for
QAOA-Max-Cut for \(p> 1\) in terms of an approximation ratio of 75.59\%. Recent approaches solve Max-Cut with coherent
networks~\cite{harrison_2022}, and also go beyond standard quantum hardware. A comprehensive evaluation
of quantum annealing performance on different QPUs for the Max-Cut problem is provided by
Willsch~\etal~\cite{willsch_2020}.

More general investigations of QAOA are plentiful; Barthi~\etal~\cite{bharti_2022} summarise
many of them. For instance, Xue~\etal~\cite{xue_2021} consider the effect of noise on QAOA
performance, while Yu~\etal~\cite{yu_2022} provide an automatic depth optimisation technique.
At present, a widely accepted and empirically observed, yet not fully understood hypothesis claims a
concentration of the optimisation parameters \(\vec{\beta}, \vec{\gamma}\) on relatively low-dimensional sub-manifolds of the
possible search space. Akshay~\etal~\cite{akshay_2021} report positive results in this directions,
and Zhou~\etal~\cite{zhou_2020} give a concrete construction to benefit from this concentration to
improve the algorithmic performance of the classical component of the hybrid algorithm
(their FOURIER construction is numerically conjectured to produce quasi-optima in
time \(\bigO{\text{poly}(p)}\) instead of \(2^{\bigO{p}}\) for the standard QAOA algorithm).
Additionally, several extensions to the QAOA algorithm have been proposed, for instance for quantum
alternating operator ansatz by Hadfield~\etal~\cite{hadfield_2019}, or the mixer-phaser ansätze by
LaRose~\etal~\cite{larose_2022}.

However, to the best of our knowledge, we are not aware of any discussion on how to improve
performance and feasibility of Max-Cut on quantum computers (QCs) using
co-design, and how to holistically evaluate potential performance benefits including the
overhead by unavoidable classical computing components beyond the optimisation algorithms
employed in the hybrid approach. We discuss these issues in this paper, and believe they hold
potential for more general insights on how to achieve first practical speedups for quantum
algorithms given realistic systemic constraints and boundary conditions.


\subsection{Modelling Max-Cut as QUBO}\label{subsec:maxcut_qaoa_basics}
Following the seminal collection of transformations provided by Lucas~\cite{lucas_2014},
the Max-Cut problem can be cast as a QUBO using binary variables $x_i$ with $x_i = 1$ if node $i$ lies within the first subset,
and $x_i=0$ if it lies in the second subset. If an edge connecting the nodes $i$ and $j$ is
part of the cut, thus connecting the two subsets, exactly one of $x_i$ and $x_j$ is equal to
zero and the other one is equal to one. In this case, $H_{i,j}=(x_i + x_j - 2 x_i x_j)$ equals
one and in the cases where $x_i = x_j$ it equals zero. Finding the maximum cut corresponds to
maximising the sum of $H_{i,j}$ over all edges of the underlying graph, or, equivalently,
minimising the sum over $-H_{i,j}$. In the following, the latter approach will be used. Thus,
the optimal solution of the Max-Cut problem can be encoded as the ground state of the Hamiltonian \(H_P = \sum_{i,j \in E} (2 x_i x_j - x_i - x_j)\),
%
%
which is passed to the generic QAOA algorithm as objective function to determine a minimum solution. 

\subsection{Setup}\label{subsec:setup}

The problem graphs \(G=(V,E)\) subjected to Max-Cut are characterised by the number of
nodes, $N =|V|$, and the graph density defined as $d=|E| / |E_{\text{max}}| \in [0, 1]$,
where $|E|$ is the number of edges of \(G\), and $|E_{\text{max}}| = n(n-1)/2$ is the
number of edges of a clique comprising \(|V|\) nodes, which upper bounds the possible
number of edges in \(G\). Each node is represented by one qubit. 

To run a quantum circuit on a QPU, it has to be compiled to meet the
requirements of the hardware. The process of compilation consists of
several steps and accounts for the limited connectivity between the
qubits as well as for the native gate set, which describes the set of
gates that can be executed on the specific hardware. Both properties
depend on the chosen technology. For instance, quantum devices
such as ion traps exhibit all-to-all connectivity, whereas others
do not. If the circuit contains gates which are not part of the native
gate set, they are decomposed accordingly. Missing connections between
qubits are countered by adding so-called SWAP-gates, which themselves
often need to be decomposed again into $\CXgate$ gates, for instance when using the IBM-Q gate set considered here. All of these steps increase the depth of the circuit. 
Within this work, creation, compilation and simulation of quantum circuits was implemented using Qiskit. For compilation (which is called transpilation in Qiskit), we use a
predefined routine of Qiskit. It consists of the following steps:
\begin{enumerate}
\item Virtual circuit optimisation, for instance, parallelisation of gates
\item Decomposition of gates containing three qubits or more into two-qubit gates
\item Placement of the virtual qubits on the physical qubits
\item Routing on coupling map, introduction of necessary SWAP gates 
\item Translation to native gate set
\item Optimisation of the resulting physical circuit
\end{enumerate}
Circuit optimisation can be executed at different levels. For all investigations in this work, the optimisation level was set to the maximal value of $3$, which describes heavy optimisation, including also re-synthesis of two-qubit blocks. The placement of the SWAP gates is performed using a stochastic method, called stochastic SWAP, which leads to different compilation results for the same initial circuit. Therefore, we repeat the compilation process $20$ times for each parameter set and consider the mean of the circuit depth over these values. 

The base coupling map is derived from the IBM-Q Washington backend with 127
qubits by adding two connections which are missing in the original hardware.
The qubits are arranged in the so-called heavy-hex lattice geometry illustrated
in Fig.~\ref{subfig:IBMQ_65}. The native gate set of IBM-Q hardware is used:
Rotation $R_{\text{Z}}$, phase shift $S_{\text{X}}$, Pauli (Not) $X$, and controlled $X$ (\CXgate). 


NISQ QPUs suffer from limited gate fidelities and noise. These effects could
also be included when simulating and compiling quantum circuits. Within the
compilation process, the placement of the virtual qubits could be adapted to the
differences in noise characteristics between the qubits, which occur for superconducting
quantum devices, minimising the overall effect of noise. However, the limited 
available space does not allow us to consider these aspects.

\section{Hardware-System Co-Design} 

\subsection{Optimisation Potentials}

QPUs require integration with classical computers to solve problems~\cite{Schoenberger_2022},
regardless if hybrid or
\enquote{pure} quantum algorithms are used. This, invariably, induces temporal overheads 
that are usually ignored when studying the complexity-theoretic performance of quantum
algorithms. However, especially for NISQ devices that are unlikely to produce exponential
speedups, such details cannot be ignored to judge potential gains by quantum
technology~\cite{Franz:2022}.
The time required to execute an algorithm on a QPU that interacts with a CPU comprises
several contributions: (a) Circuit execution time \tci{} on the QPU (also
considering the number \nsa{} of samples required to obtain accurate statistics),
(b) time \tm{} for performing measurements of quantum states, (c) classical
parameter optimisation time \topt{} on the CPU, (d) the amount of 
optimisation iterations \nit{}, and (e) time for passing input parameters
and output results between QPU and CPU \tco.  A straightforward model to describe
the execution time required for hybrid algorithms like QAOA is therefore given by
\begin{equation}
    T = \nit\cdot[\nsa\cdot(\tci+\tm) + \topt + \tco].\label{eq:runtime}
\end{equation}

\subsection{Parameter Estimation on IBM-Q hardware}
\newcommand{\mus}{\ensuremath{\mu\text{s}}}

To evaluate optimisation potentials using the model in Eq.~(\ref{eq:runtime}), we
need to determine performance values on QPUs. 
Obviously, these parameters not only depend on size and possibly structure of the input instance, but also on the underlying execution platform. 
To illustrate the \emph{relative} influence of the factors for a typically sized
instance, it suffices to consider one set of parameters, for which we chose a
graph with 20 nodes (amounting to 20 qubits) and a graph density of 0.5 solved using QAOA with a single layer.

In the following, we estimate the runtime of such a QAOA circuit
using a custom routine that pre-compiles the logical 
circuit to the gate set provided by IBM-Q hardware,
and then executes the result on a simulator. We deliberately do not resort
to physical hardware to avoid any degradation by noise, as our goal is to find
optimisation potentials for QPU-system co-design, not to evaluate limitations of current NISQ devices. The QAOA circuit is
parameterised by $\vec{\beta}$ and $\vec{\gamma}$, which are optimised using the COBYLA
routine provided by Qiskit. As an objective function, we compute the mean expectation value of $H_P$, as described in Sec.~\ref{subsec:maxcut_qaoa_basics}. (see the \href{https://github.com/lfd/arcs2022.git}{replication package} for details). We sample the quantum circuit using $ \nsa = 1,024$ shots in each iteration. We find \(\topt=159\mus\), \(\tm=28.6\mus\), and \(\nit=25\).

A quantum circuit comprises different
gates, and the time it takes each gate to operate is denoted as \emph{gate time}.
To obtain an estimate of the distribution of classical and quantum contributions to
the total algorithmic runtime, we estimate the quantum circuit execution time $\tci$ from the (known) hardware gate times
and the structure of the circuit. 

 \begin{wraptable}[11]{r}{0.35\textwidth}
 \vspace*{-2.5em}
    \caption{Gate times from backend FakeBrooklyn.}
    \vspace*{-1em}
	\label{tbl:gate_times}
	\begin{center}\begin{tabular}{l@{\extracolsep{1em}}r@{\extracolsep{1em}}r}
		\toprule
		Gate & Exec\(^{\text{\underline{n}}}\) & Std.\\
		     & time [ns] & Dev.\\
		\midrule
			$R_Z$ & 0 & 0\\
			$S_X$ & 35.56 & 0 \\
			$X$ & 35.56 & 0\\
			$\CXgate$ & 370 & 80\\
    \bottomrule
	\end{tabular}\end{center}
	\vspace*{-3em}
 \end{wraptable}

Approximate values for the gate times as provided by the Qiskit \emph{mock backend} FakeBrooklyn are used.
In case of \CXgate gates, different values arise for different possible qubit
pairs, and we consider the average value. The execution times for the single
qubit gates are identical for all qubits. Table~\ref{tbl:gate_times} lists the
concrete gate times. The execution time of the circuit calculated from the gate
times is \(\tci=120 \pm 20\mus\), averaged over $20$ circuit transpilation runs. 

\begin{wraptable}[11]{l}{0.4\textwidth}
	\vspace*{-1.5em}\caption{Communication and total execution time \(T\) for
	QPU deployment scenarios.}\label{tbl:deployment}
    \centering\begin{tabular}{lr@{\extracolsep{1em}}r}
        \toprule
        Scenario & \(\tco\) & \(T\)\\
        \midrule
        Cloud & 50ms & 5.07s\\
        Local Bus & 1ms & 3.84s\\
        SoC\footnote{SoC denotes a system-on-chip solution, where CPU and QPU are integrated on the chip level.} & 25\(\mu\text{s}\) & 3.82s\\
    \bottomrule
    \end{tabular}
\end{wraptable}

The CPU-QPU communication time \tco{} depends on how QPUs are deployed,
and may vary over several orders of magnitude.
We consider three different scenarios in Table~\ref{tbl:deployment}: (a) access
to a QPU via cloud services, with a communication round-trip
time of about 50ms; (b) a QPU with direct attachment
to the local CPU (for instance via direct LAN connection), for which we
expect \(\tco\approx 1\text{ms}\), and an integrated system comprising a QPU and CPU that communicate
via an internal system bus with \(\tco\approx 25\mu\text{s}\).\footnote{The given communication times are rough estimates supposed to \emph{illustrate} 
optimisation potentials and relative parameter influence. We obtained the
numbers by measuring typical \texttt{ping} durations in cloud and local network
scenarios, and estimate QPU-CPU communication time by the round-trip time of
an inter-processor-interrupt in a RiscV-system, given the assumptions that 
a QPU-CPU SoC design will likely be based on modifiable classical architectures, 
and that communication times between QPU and CPU are similar to inter-core
communication times.\newline
Owing to the restricted space, we do not provide more fine-grained and realistic
estimates of and models for these quantities, but remark that they would very
likely not substantially change our findings and conclusions.} 
Assume that, as in many scenarios of practical interest, we deal with a
\enquote{typically} sized problem instance, for which we can assume that
\(\nit\), \(\nsa\) and \(\topt\) are approximately independent of the
specific input. This leaves the communication time \(\tco\) and the circuit
time \(\tci\) as candidates for optimisation. With \(\nit\), \(\nsa\) and \(\topt\)
constant, they linearly contribute to the growth of the total execution time
\(T\), and the corresponding slopes are 25 for \(\tco\) and 25,600 for \(\tci\).
The magnitude of the latter slope would seem to indicate that reducing
communication times holds greater optimisation potential than decreasing
circuit execution times. Alas, as Table~\ref{tbl:deployment} indicates,
moving from a local QPU deployment to on-chip integration is much less
beneficial than moving from a cloud deployment to local operation of
QPUs.\footnote{Moving from locally connected components to on-chip integration
might be beneficial for latency-critical embedded systems with quantum acceleration,
but is likely not overly relevant for many HPC use-cases.}

However, when we consider the \enquote{Local Bus} scenario and assume that
the circuit execution time can be reduced from $120\mu\text{s}$ to $30\mu\text{s}$ 
(see Fig.~\ref{fig:runtimes} and later explanations for a rationale), the overall execution time reduces from \(T=3.84\text{s}\) to
\(T=1.28\text{s}\), which indicates substantial potential for optimisation using
hardware adaption.
Comparing the baseline scenario (cloud + standard topology) with the co-design
results (local communication, adapted topology), we find a total execution time
reduction of \(1-1.28/5.07\approx 75\%\), which may substantially impact practical
scenarios, especially given that accelerators often solve the same primitive
repeatedly in inner loops.

\subsection{Physical Possibilities and Limitations}
Different properties of QPUs can be considered to reduce circuit execution times.
On the one hand, there is the geometric layout of the qubits and their connectivity.
On the other hand, the native gate set, as well as the fidelities and execution times
of the gates influence the performance of algorithms, as well as the effects of noise.
Depending on the hardware platform, the gate times and fidelities can differ for
each individual (pair of) qubit(s), as it is the case for superconducting qubits. 
In this work, we focus on the effects of the qubit connectivity and the number of qubits available on the device compared to the problem size.


Changing the qubit connectivity for superconducting devices necessitates to physically re-wire qubits. Moreover, issues such as cross-talk can occur if the connectivity
increases too much. This is reflected in the heavy-hex lattice design of 
IBM-Q, which features reduced connectivity compared to previous layouts.

In contrast, trapped ion quantum computers feature all-to-all
connectivity~\cite{bruzewic_2019}, but are currently limited to few qubits
($\approx 20$). QCs based on neutral atom technology do not feature all-to-all
connectivity, but their connectivity is usually higher than nearest-neighbour,
and can be further increased. 

\subsection{Variation of the Coupling Density}

The connectivity between qubits is described by the coupling density $c$
given by \(c = N_C/N_{C, \mathrm{max}}\), where $N_C$ denotes the number of
connections between pairs of qubits (that is, the possible interactions), and $N_{C, \mathrm{max}}$ gives the maximal \emph{possible} number of connections, which is obviously reached for a clique connectivity with \(N_{C, \mathrm{max}} = n(n-1)/2\) for \(n\) available qubits. Thus, $c=1$ describes a quantum device with all-to-all connectivity. The base topology of the IBM-Q devices has a coupling density of $c \approx 0.0139$. For the simulations in this subsection, the size of the backend is kept constant at 127 qubits. The coupling density is increased by randomly adding connections between the qubits.\footnote{The placement of new connection favours augmenting regions with existing high connectivity density, following the assumption that adding extra connections is easier for regions that are already well connected. Given the lack of space, we refer readers to the \href{https://github.com/lfd/arcs2022.git}{replication package} for the exact details.} Each data point shown in this section is an average over $20$ compilation runs, using again the standard transpilation process of Qiskit with optimisation level 3.




\vspace*{-1em}
\begin{figure*}[htb]
  \begin{center}
    \input{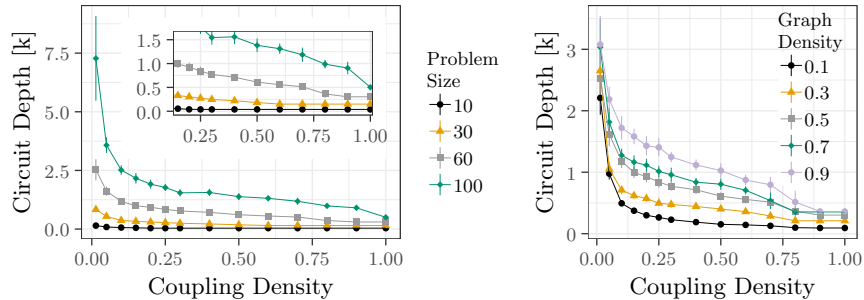}\vspace*{-1em}
    \caption{Mean circuit depth vs. coupling density for $p=1$ (127
      qubit backend), varying problem size (lhs) for $d=0.5$ and
      graph density (rhs) for $N=60$.}\label{fig:depth_vs_cmap_density}
  \end{center}
\end{figure*}
\vspace*{-1.25em}

We investigate the effect of increasing coupling density for different problem sizes,
graph densities, and QAOA layers. In general, a higher coupling density reduces the number
of SWAP gates needed to realise the desired two-qubit interactions, and decreases
circuit depth.
This is evident in the left panel of Fig.~\ref{fig:depth_vs_cmap_density}, where the resulting circuit depth is plotted vs.\ coupling density for different problem sizes and $d=0.5$. The
circuit depth saturates for higher coupling densities.
The saturation density $c_{\mathrm{sat}}$ as well as the saturation value of the circuit depth increase with the problem size, which can also be seen in the inset of the plot. Solving larger problem instances requires more qubits and thus also more two-qubit gates, which leads to deeper circuits in general.

\begin{wrapfigure}[10]{l}{0.4\linewidth} 
\vspace{-3.5em}
  \begin{center}
\begin{tikzpicture}[x=1pt,y=1pt]
\definecolor{fillColor}{RGB}{255,255,255}
\path[use as bounding box,fill=fillColor,fill opacity=0.00] (0,0) rectangle (130.17, 95.46);
\begin{scope}
\path[clip] (  0.00,  0.00) rectangle (130.17, 95.46);
\definecolor{drawColor}{RGB}{255,255,255}
\definecolor{fillColor}{RGB}{255,255,255}

\path[draw=drawColor,line width= 0.5pt,line join=round,line cap=round,fill=fillColor] (  0.00, -0.00) rectangle (130.17, 95.46);
\end{scope}
\begin{scope}
\path[clip] ( 32.83, 27.90) rectangle (125.17, 90.46);
\definecolor{fillColor}{RGB}{255,255,255}

\path[fill=fillColor] ( 32.83, 27.90) rectangle (125.17, 90.46);
\definecolor{drawColor}{gray}{0.92}

\path[draw=drawColor,line width= 0.3pt,line join=round] ( 32.83, 34.06) --
	(125.17, 34.06);

\path[draw=drawColor,line width= 0.3pt,line join=round] ( 32.83, 59.45) --
	(125.17, 59.45);

\path[draw=drawColor,line width= 0.3pt,line join=round] ( 32.83, 84.84) --
	(125.17, 84.84);

\path[draw=drawColor,line width= 0.3pt,line join=round] ( 46.48, 27.90) --
	( 46.48, 90.46);

\path[draw=drawColor,line width= 0.3pt,line join=round] ( 67.77, 27.90) --
	( 67.77, 90.46);

\path[draw=drawColor,line width= 0.3pt,line join=round] ( 89.05, 27.90) --
	( 89.05, 90.46);

\path[draw=drawColor,line width= 0.3pt,line join=round] (110.33, 27.90) --
	(110.33, 90.46);

\path[draw=drawColor,line width= 0.5pt,line join=round] ( 32.83, 46.75) --
	(125.17, 46.75);

\path[draw=drawColor,line width= 0.5pt,line join=round] ( 32.83, 72.15) --
	(125.17, 72.15);

\path[draw=drawColor,line width= 0.5pt,line join=round] ( 35.84, 27.90) --
	( 35.84, 90.46);

\path[draw=drawColor,line width= 0.5pt,line join=round] ( 57.13, 27.90) --
	( 57.13, 90.46);

\path[draw=drawColor,line width= 0.5pt,line join=round] ( 78.41, 27.90) --
	( 78.41, 90.46);

\path[draw=drawColor,line width= 0.5pt,line join=round] ( 99.69, 27.90) --
	( 99.69, 90.46);

\path[draw=drawColor,line width= 0.5pt,line join=round] (120.97, 27.90) --
	(120.97, 90.46);
\definecolor{drawColor}{RGB}{0,0,0}

\path[draw=drawColor,line width= 0.2pt,line join=round] ( 37.03, 66.76) -- ( 37.03, 87.62);

\path[draw=drawColor,line width= 0.2pt,line join=round] ( 40.10, 52.85) -- ( 40.10, 59.68);

\path[draw=drawColor,line width= 0.2pt,line join=round] ( 44.36, 44.48) -- ( 44.36, 49.71);

\path[draw=drawColor,line width= 0.2pt,line join=round] ( 48.61, 40.86) -- ( 48.61, 46.21);

\path[draw=drawColor,line width= 0.2pt,line join=round] ( 52.87, 36.85) -- ( 52.87, 43.39);

\path[draw=drawColor,line width= 0.2pt,line join=round] ( 57.13, 36.39) -- ( 57.13, 41.40);

\path[draw=drawColor,line width= 0.2pt,line join=round] ( 61.38, 34.65) -- ( 61.38, 38.53);

\path[draw=drawColor,line width= 0.2pt,line join=round] ( 69.90, 30.74) -- ( 69.90, 33.95);

\path[draw=drawColor,line width= 0.2pt,line join=round] ( 78.41, 31.97) -- ( 78.41, 31.99);

\path[draw=drawColor,line width= 0.2pt,line join=round] ( 86.92, 31.97) -- ( 86.92, 31.99);

\path[draw=drawColor,line width= 0.2pt,line join=round] ( 95.43, 31.97) -- ( 95.43, 31.99);

\path[draw=drawColor,line width= 0.2pt,line join=round] (103.95, 31.99) -- (103.95, 31.99);

\path[draw=drawColor,line width= 0.2pt,line join=round] (112.46, 31.99) -- (112.46, 31.99);

\path[draw=drawColor,line width= 0.2pt,line join=round] (120.97, 31.99) -- (120.97, 31.99);
\definecolor{fillColor}{RGB}{0,0,0}

\path[draw=drawColor,line width= 0.8pt,line join=round,line cap=round,fill=fillColor] ( 37.03, 77.19) circle (  1.35);

\path[draw=drawColor,line width= 0.8pt,line join=round,line cap=round,fill=fillColor] ( 40.10, 56.26) circle (  1.35);

\path[draw=drawColor,line width= 0.8pt,line join=round,line cap=round,fill=fillColor] ( 44.36, 47.10) circle (  1.35);

\path[draw=drawColor,line width= 0.8pt,line join=round,line cap=round,fill=fillColor] ( 48.61, 43.54) circle (  1.35);

\path[draw=drawColor,line width= 0.8pt,line join=round,line cap=round,fill=fillColor] ( 52.87, 40.12) circle (  1.35);

\path[draw=drawColor,line width= 0.8pt,line join=round,line cap=round,fill=fillColor] ( 57.13, 38.89) circle (  1.35);

\path[draw=drawColor,line width= 0.8pt,line join=round,line cap=round,fill=fillColor] ( 61.38, 36.59) circle (  1.35);

\path[draw=drawColor,line width= 0.8pt,line join=round,line cap=round,fill=fillColor] ( 69.90, 32.35) circle (  1.35);

\path[draw=drawColor,line width= 0.8pt,line join=round,line cap=round,fill=fillColor] ( 78.41, 31.98) circle (  1.35);

\path[draw=drawColor,line width= 0.8pt,line join=round,line cap=round,fill=fillColor] ( 86.92, 31.98) circle (  1.35);

\path[draw=drawColor,line width= 0.8pt,line join=round,line cap=round,fill=fillColor] ( 95.43, 31.98) circle (  1.35);

\path[draw=drawColor,line width= 0.8pt,line join=round,line cap=round,fill=fillColor] (103.95, 31.99) circle (  1.35);

\path[draw=drawColor,line width= 0.8pt,line join=round,line cap=round,fill=fillColor] (112.46, 31.99) circle (  1.35);

\path[draw=drawColor,line width= 0.8pt,line join=round,line cap=round,fill=fillColor] (120.97, 31.99) circle (  1.35);

\path[draw=drawColor,line width= 0.2pt,line join=round] ( 37.03, 77.19) --
	( 40.10, 56.26) --
	( 44.36, 47.10) --
	( 48.61, 43.54) --
	( 52.87, 40.12) --
	( 57.13, 38.89) --
	( 61.38, 36.59) --
	( 69.90, 32.35) --
	( 78.41, 31.98) --
	( 86.92, 31.98) --
	( 95.43, 31.98) --
	(103.95, 31.99) --
	(112.46, 31.99) --
	(120.97, 31.99);
\definecolor{drawColor}{gray}{0.20}

\path[draw=drawColor,line width= 0.5pt,line join=round,line cap=round] ( 32.83, 27.90) rectangle (125.17, 90.46);
\end{scope}
\begin{scope}
\path[clip] (  0.00,  0.00) rectangle (130.17, 95.46);
\definecolor{drawColor}{gray}{0.30}

\node[text=drawColor,anchor=base east,inner sep=0pt, outer sep=0pt, scale=  0.80] at ( 28.33, 44.00) {50};

\node[text=drawColor,anchor=base east,inner sep=0pt, outer sep=0pt, scale=  0.80] at ( 28.33, 69.39) {100};
\end{scope}
\begin{scope}
\path[clip] (  0.00,  0.00) rectangle (130.17, 95.46);
\definecolor{drawColor}{gray}{0.20}

\path[draw=drawColor,line width= 0.5pt,line join=round] ( 30.33, 46.75) --
	( 32.83, 46.75);

\path[draw=drawColor,line width= 0.5pt,line join=round] ( 30.33, 72.15) --
	( 32.83, 72.15);
\end{scope}
\begin{scope}
\path[clip] (  0.00,  0.00) rectangle (130.17, 95.46);
\definecolor{drawColor}{gray}{0.20}

\path[draw=drawColor,line width= 0.5pt,line join=round] ( 35.84, 25.40) --
	( 35.84, 27.90);

\path[draw=drawColor,line width= 0.5pt,line join=round] ( 57.13, 25.40) --
	( 57.13, 27.90);

\path[draw=drawColor,line width= 0.5pt,line join=round] ( 78.41, 25.40) --
	( 78.41, 27.90);

\path[draw=drawColor,line width= 0.5pt,line join=round] ( 99.69, 25.40) --
	( 99.69, 27.90);

\path[draw=drawColor,line width= 0.5pt,line join=round] (120.97, 25.40) --
	(120.97, 27.90);
\end{scope}
\begin{scope}
\path[clip] (  0.00,  0.00) rectangle (130.17, 95.46);
\definecolor{drawColor}{gray}{0.30}

\node[text=drawColor,anchor=base,inner sep=0pt, outer sep=0pt, scale=  0.80] at ( 35.84, 17.89) {0.00};

\node[text=drawColor,anchor=base,inner sep=0pt, outer sep=0pt, scale=  0.80] at ( 57.13, 17.89) {0.25};

\node[text=drawColor,anchor=base,inner sep=0pt, outer sep=0pt, scale=  0.80] at ( 78.41, 17.89) {0.50};

\node[text=drawColor,anchor=base,inner sep=0pt, outer sep=0pt, scale=  0.80] at ( 99.69, 17.89) {0.75};

\node[text=drawColor,anchor=base,inner sep=0pt, outer sep=0pt, scale=  0.80] at (120.97, 17.89) {1.00};
\end{scope}
\begin{scope}
\path[clip] (  0.00,  0.00) rectangle (130.17, 95.46);
\definecolor{drawColor}{RGB}{0,0,0}

\node[text=drawColor,anchor=base,inner sep=0pt, outer sep=0pt, scale=  1.00] at ( 79.00,  6.94) {Coupling Density};
\end{scope}
\begin{scope}
\path[clip] (  0.00,  0.00) rectangle (130.17, 95.46);
\definecolor{drawColor}{RGB}{0,0,0}

\node[text=drawColor,rotate= 90.00,anchor=base,inner sep=0pt, outer sep=0pt, scale=  1.00] at ( 11.89, 59.18) {Runtime [us]};
\end{scope}
\end{tikzpicture}\vspace*{-1em}
    \caption{Circuit runtime for \(N=20, d=0.5, p=1\).}
    \label{fig:runtimes}
  \end{center}
\end{wrapfigure}
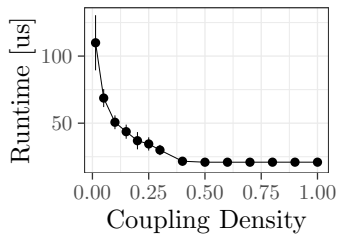
Modifying  the graph density for a fixed problem size does not change $c_{\mathrm{sat}}$,
as illustrated by the data in the right panel of Fig.~\ref{fig:depth_vs_cmap_density} for $N=60$, where
the circuit depth remains constant for $c > 0.8$. The overall circuit depth and its
saturation value increase with graph density. This can be traced back to 
an increased density of the QUBO resulting from an increasing amount of edges that
necessitate more two-qubit gates. For larger number of QAOA
layers, the circuit depth increases linearly, whereas the saturation density
$c_{\mathrm{sat}}$ remains unchanged. 
The decrease in circuit depth for higher coupling densities results in shorter
runtimes as illustrated in Fig.~\ref{fig:runtimes}. Since a relatively small problem
instance with $20$ qubits is considered, the runtime saturates for moderate coupling
densities, similar as the circuit depth (see Fig.~\ref{fig:depth_vs_cmap_density}). 

Both graphs in Fig.~\ref{fig:depth_vs_cmap_density} show an important trend: Even a moderate increase in coupling
density causes a substantial decrease in circuit depth - growing the coupling
density from the standard topology to a 10\% extended density reduces circuit depth
for \(N=100\) from 7,000 to slightly over 2,000 (the effect is similar, yet becomes
less pronounced for small input instances). We find this decrease to be a crucial improvement
with regards to circuit execution times, but it also benefits NISQ systems, since
shorter circuits pick up less effects of noise. Given limited space, we can 
unfortunately not study the impact of noise further in this paper, but retain this aspect for future work.

\vspace*{-1em}
\begin{figure*}[hbp]
  \input{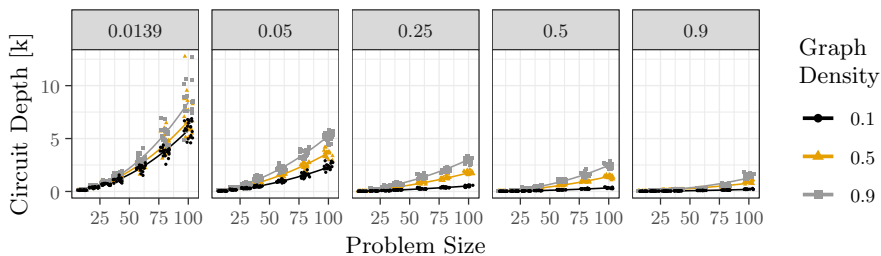}\vspace*{-2em}
  \caption{Circuit depth growth behaviour over problem size for
  different coupling densities in the panels. Solid lines represent regression
  models for varying input graph dependencies.}\label{fig:model_fits}
\end{figure*}
\vspace*{-1em}

To quantify the growth of the circuit depth with problem size, we construct a univariate regression model\footnote{Technically, we employ a robust quantile regression approach~\cite{koenker_2022} because 
the stochastic circuit generation process produces pronounced
outliers.} of the form \(f(x) = c_{0} + c_{1}x + c_{2}x^{2}\).
We have performed the usual regression diagnostics, and an ANOVA based model
selection procedure unambiguously confirms that the choice of quadratic growth behaviour is
preferable to linear and exponential alternative models. Since Fig.~\ref{fig:model_fits}
visually demonstrates an excellent match between data and model, we do not explicitly 
spell out details of these diagnostics.

\begin{figure*}[htb]
  \input{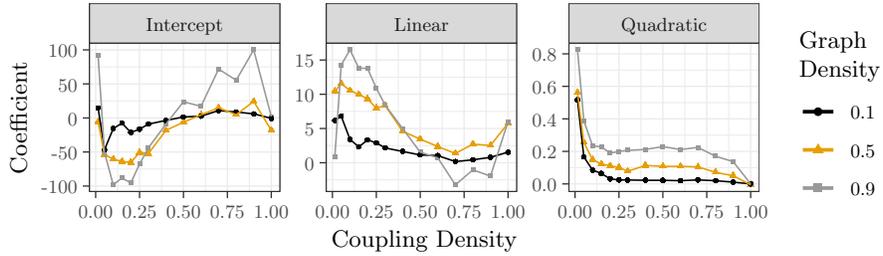}\vspace*{-2em}
  \caption{Coefficients of the quadratic quantile regression model depending on graph and coupling density.}
  \label{fig:regression_coefficients}
\end{figure*}

It is interesting to observe the behaviour of the regression coefficients 
with increasing connectivity map density in Fig.~\ref{fig:regression_coefficients}:
The quadratic contribution is most pronounced for the unmodified topology, but quickly
wanes with increasing connectivity map densities, and saturates in the connectivity 
density map range \([0.25, 0.75]\). Regardless of input problem structure (graph
density), any quadratic contribution to growth vanishes for fully connected topologies.
In general, more effort in implementing physical connections pays off with 
more favourable QPU scaling behaviour.
\vspace*{-1em}
\vspace*{-1em}

\subsection{Variation of the Backend Size}

 \begin{wraptable}[11]{r}{0.65\linewidth}
	\vspace*{-2.5em}\caption{Backend sizes used for the data in Fig.~\ref{fig:ext_top_depth_vs_size_N60_d0.5}, characterised by the number of unit cell rows $N_{\mathrm{rows}}$, unit cell columns $N_{\mathrm{cols}}$ and number of qubits.}
	\label{tbl:backend_sizes}
	\begin{tabular}{lll|lll}
   \toprule%
		$N_{\mathrm{rows}}$ \; & $N_{\mathrm{cols}}$ \; & $N_{\mathrm{qubits}}$ \;\;\; & $N_{\mathrm{rows}}$ \; & $N_{\mathrm{cols}}$ \; & $N_{\mathrm{qubits}}$ \; \\
		\toprule
			4 & 2 & 65 & 5 & 3 & 108\\
			3 & 3 & 70 & 4 & 4 & 113\\
			5 & 2 & 79 & 6 & 3 & 127\\
			4 & 3 & 89 & 5 & 4 & 137\\
			6 & 2 & 93 & 6 & 4 & 161\\	
	\bottomrule
	\end{tabular}
 \end{wraptable}
 The properties of the compiled circuit also depend on the size of the quantum device, namely
 the number of available qubits relative to the problem size. If the backend has more qubits
 available than needed, there are more possibilities for routing the virtual to physical
 qubits. In general, a more efficient placement of the circuit on the hardware can be found,
 for instance, by assigning more of the virtual qubit pairs that share two-qubit gates to
 physical qubits which are directly connected. 

\begin{figure}[htb]
\centering
 \begin{subfigure}[t]{0.25\linewidth}
 \centering
   \includegraphics[width=\linewidth]{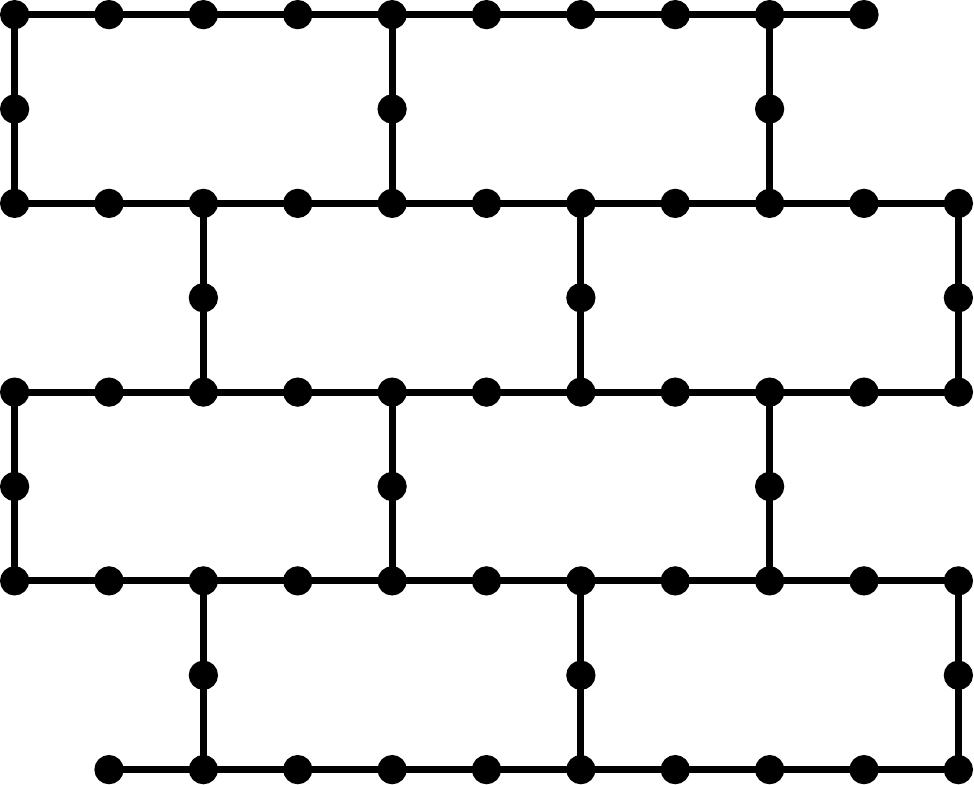}
   \caption{$c\approx 0.0139$}
   \label{subfig:IBMQ_65}
 \end{subfigure}
 \hspace{\fill}
 \begin{subfigure}[t]{0.25\linewidth}
  \centering
   \includegraphics[width=\linewidth]{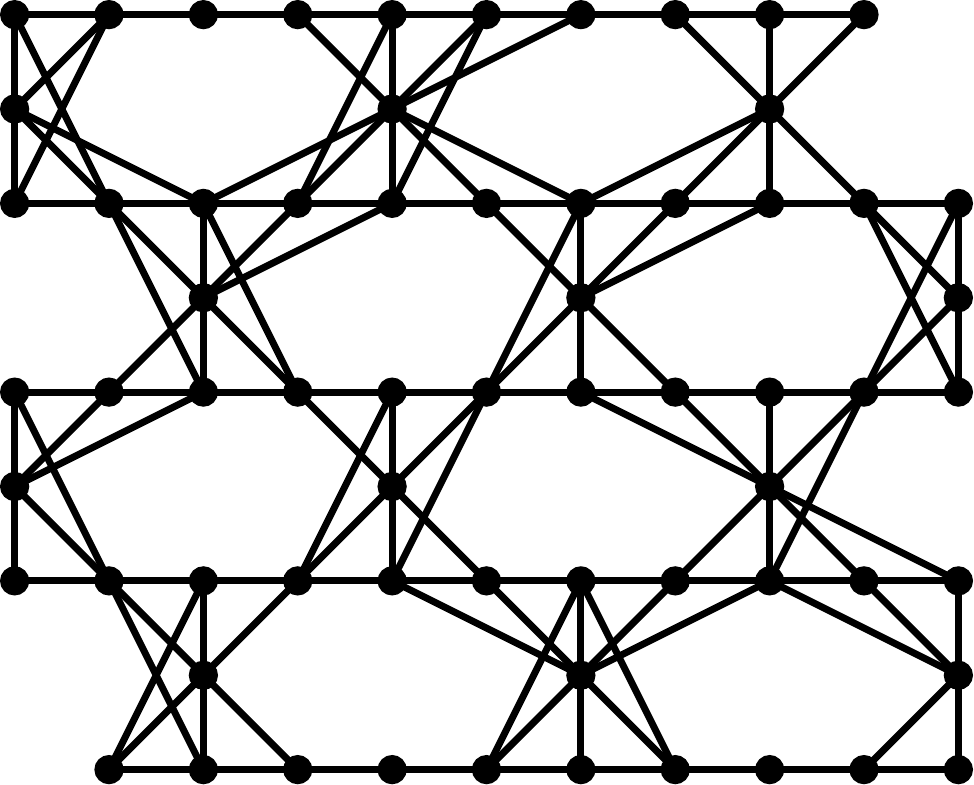}
      \caption{$c = 0.1$}
      \label{subfig:IBMQ_65_c_0.1}
  \end{subfigure}
   \hspace{\fill}
 \begin{subfigure}[t]{0.25\linewidth}
  \centering
   \includegraphics[width=\linewidth]{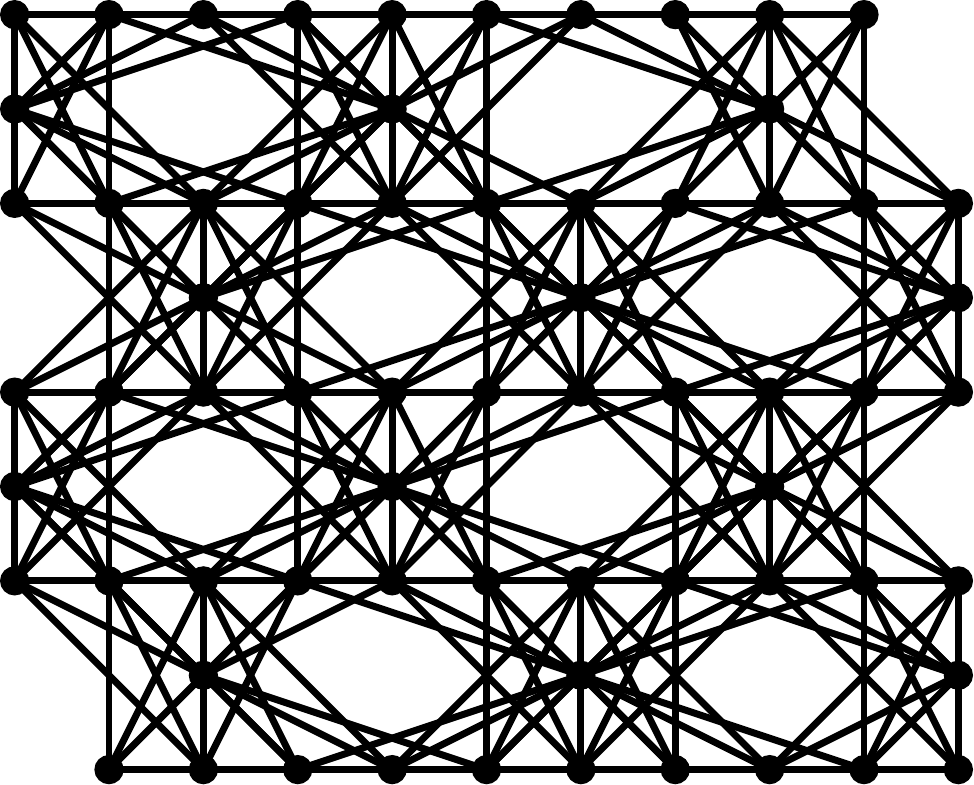}
      \caption{$c = 0.2$}
      \label{subfig:IBMQ_65_c_0.2}
  \end{subfigure}
\caption{Sketch of the heavy-hex lattice coupling map for $65$ qubits with base (a) and
extended (b, c) coupling density.}
\label{fig:IBMQ_change_cmap_density}
\end{figure}
\vspace*{-1em}

To examine the influence of the backend size on the circuit depth, the circuit for
a Max-Cut problem with $N=60$, $d=0.5$ and $p=1$ is compiled on backends of different
sizes between $65$ and $161$ qubits. The qubits are arranged in the heavy-hex lattice
geometry and the size is increased by successively adding unit cells below or on the
right. The smallest backend corresponds to the IBM-Q Brooklyn device, consisting of
four rows and two columns of unit cells, as depicted in
Fig.~\ref{fig:IBMQ_change_cmap_density} for the base coupling density as well as $c=0.1$ and $c=0.2$.
Starting from this layout, ten different sizes are considered, as summarised in Tab.~\ref{tbl:backend_sizes}. 

\begin{wrapfigure}[13]{r}{0.45\linewidth} 
\vspace{-3.5em}
\begin{center}
    \input{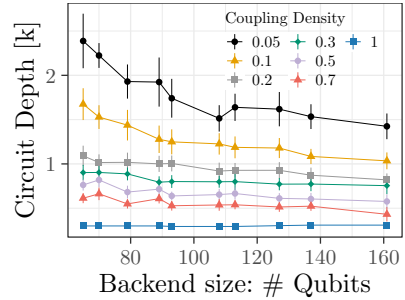}\vspace*{-2em}
    \caption{Mean depth vs. backend size for $N=60$, $d=0.5$ and $p=1$.}\label{fig:ext_top_depth_vs_size_N60_d0.5}
\end{center}
\end{wrapfigure}

In Fig.~\ref{fig:ext_top_depth_vs_size_N60_d0.5}, the resulting mean
circuit depths are shown as a function of the number of qubits for
various coupling densities. In general, the circuit depth decreases
with the backend size, illustrating the effects of a more efficient
placement of the circuit described above. These effects become less
pronounced for higher coupling densities, since the backend then
exhibits more qubit pairs that are directly connected. Consequently,
for a backend with all-to-all connectivity ($c=1$), the backend size
has no influence on the depth of the compiled circuit.

\section{Conclusion \& Outlook} 
In this paper, we have investigated integration and co-design possibilities for QPUs
that are supposed to act as computational accelerators in high performance computing 
systems. Our results show that designing purpose-specific QPUs with adapted 
topologies holds promises in terms of computational capabilities. We have
also shown that circuit depth is reduced by increasing the connectivity map density, but 
already saturates at values $c \ll 1$, with a slight dependency on the problem size. Thus,
all-to-all connectivity is not needed in all cases, relaxing the requirements on the
quantum hardware devices. We have also discussed that how integration of
QPUs is performed can have effects on their capabilities, even
if these are not as pronounced as for topology adaptations.

For now, we have focused on perfect QPUs that do not suffer from noise and imperfections.
Future work will incorporate these deficiencies into our analysis, which is important
to transfer our results to present-day NISQ systems, and will help progressing towards 
practical utility of early-stage quantum computers.
%
%
%
%

\begin{small}
  \vspace*{0.5em}\emph{Acknowledgement:} We thank Manuel Schönberger for providing his
  topology adaptation simulation code as starting point for our
  efforts.
\end{small}

\bibliographystyle{IEEEtran}
\bibliography{ARCS_2022_paper_40}
\end{document}